\documentclass[aps,twocolumn,pra]{revtex4-1}

\newcommand{\ie}{{\it i.~$\!$e.}}
\usepackage{hyperref}
\begin{document}
\title{Symplectic group methods and the Arthurs Kelly model of
 measurement in quantum mechanics}
\author{Arvind}
\email{arvind@iiser.mohali.ac.in}
\affiliation{Department of Physical Sciences, Indian
Institute of Science Education and Research (IISER) Mohali,
Sector 81 SAS Nagar, Manauli PO 140306,
Punjab, India}
\author{S. Chaturvedi}
\email{subhash@iiserb.ac.in}
\affiliation{Department of Physics, Indian Institute of
Science Education and Research (IISER) Bhopal, Bhopal Bypass
Road, Bhauri, Bhopal 462066, India}
\author{N. Mukunda}
\email{nmukunda@gmail.com}
\affiliation{Adjunct Professor, Department of Physics,
Indian Institute of Science Education and Research (IISER)
Bhopal, Bhopal Bypass Road, Bhauri,
Bhopal 462066, India }
\begin{abstract}
We study the use of methods based on the real symplectic
groups $Sp(2n,\mathcal{R})$  in the analysis of the Arthurs-Kelly
model of proposed simultaneous measurements of position and
momentum in quantum mechanics. Consistent with the fact that
such measurements are in fact not possible, we show that
the observable consequences of the Arthurs-Kelly interaction
term are contained in the symplectic transformation law
connecting the system plus apparatus variance matrices at an
initial and a final time. The individual variance matrices
are made up of averages and spreads or uncertainties for
single hermitian observables one at a time, which are quantum
mechanically well defined. The consequences of the multimode
symplectic covariant Uncertainty Principle in the 
Arthurs-Kelly context are examined.   
\end{abstract}
\maketitle
\section{Introduction}
\label{intro}
The problem of understanding the measurement process in
quantum mechanics has been of long standing interest, and
has seen a significant revival in recent
times~\cite{wigner_measurement}. The quantum Zeno
effect~\cite{sud_zeno}, the concept of weak
measurements~\cite{ahranov_weak,hari_dass_weak}, and efforts
to understand the Born Rule from more basic
principles~\cite{born_apoorva}, may be mentioned in this
context. Joint measurements of non-commuting
observables, which is the main theme of this paper, have 
been considered by several authors~\cite{ak_1965,roy_joint,bush_joint,appleby_1998}.
Weak sequential measurements of non-commuting
observables have also been considered for 
state tomography~\cite{arvind_deb_pra,arvind_deb_jpa}.

Probably the earliest model of measurement in quantum
mechanics is the one formulated by von Neumann in 1932, very
soon after the discovery of quantum mechanics itself
~\cite{vonNeumann_book}. In this model, both the system
$\mathcal{S}$ being studied and the apparatus $\mathcal{A}$
are treated quantum mechanically, and the measurement is
described by a suitable coupling term in the total
Hamiltonian. The Born Rule remains as something to be
imposed externally.

An interesting approach to the general measurement problem
is due to Sudarshan from 1976~\cite{sud_pramana_76}. Here
while the system $\mathcal{S}$ is of course quantum
mechanical, the apparatus $\mathcal{A}$ is initially
regarded as a classical system. However it is then recast as
a quantum system subject to continuous super selection
rules, and then the possibility of coupling of $\mathcal{S}$
and $\mathcal{A}$ as parts of an overall quantum system is
studied. In this approach, while the Born Rule remains
`unexplained', the well known restriction in quantum
mechanics that only mutually commuting -- \ie, compatible
-- dynamical variables can be simultaneously measured, is
derived.

As just mentioned, quantum mechanics does not permit
simultaneous measurement of non-commuting dynamical variables
such as a coordinate $\hat{q}$ and its canonically conjugate
momentum $\hat{p}$. A very interesting approach to
measurement in this situation, treating $\hat{q}$ and
$\hat{p}$ on the same footing, is the Arthurs-Kelly
($A$-$K$) proposal of 1964~\cite{ak_1965}. In the simplest
case where the system $\mathcal{S}$ is based on one
canonical pair of hermitian Cartesian operators $\hat{q}$
and $\hat{p}$, the apparatus $\mathcal{A}$ is taken to be a
quantum system involving two kinematically independent
canonical pairs of operators $\hat{Q}_1, \hat{P}_1$ and
$\hat{Q}_2, \hat{P}_2$. The idea is to use the commuting
operators $\hat{Q}_1$ and $\hat{Q}_2$ to act as pointer
positions to track the values of $\hat{q}$ and $\hat{p}$
respectively, using von Neumann type coupling terms in the
total Hamiltonian.

An interesting consequence of the $A$-$K$ model, frequently
mentioned as an important feature of it, is a kind of
uncertainty principle for the pair $\hat{Q}_1$ and
$\hat{Q}_2$, which a priori are compatible variables. It
states that the lower bound on the product of their
uncertainties is twice that for the familiar canonical
$\hat{q}-\hat{p}$ pair, paying due attention to the
differences in physical dimensions in the two cases, and
this is ascribed to inherent and unavoidable extra noise in
joint quantum measurements.

The purpose of the present work is to revisit the $A$-$K$
model, in particular to explore the use of methods based on
the real symplectic groups in this context. As has been
shown elsewhere, for quantum systems involving, say, $N$
canonical pairs of operators of the $\hat{q}-\hat{p}$ type,
the most general statement of the multimode Uncertainty
Principle is $Sp(2n, \mathcal{R})$ covariant, and is best
understood and stated using the properties of these groups.
As the original Arthurs-Kelly model involves a system with
one degree of freedom  and an apparatus with two degrees of
freedom, the relevant group here is
$Sp(6,\mathcal{R})$.

The contents of the paper are arranged as follows.
Section~\ref{ak_model} recapitulates the kinematics and
interaction term in the Arthurs-Kelly model, generalized to
have two independent coupling constants. The solution of the
Schr\"odinger equation for a general initial condition, as
well as for product initial wavefunctions, are given.
Section~\ref{sp6r} identifies the Hamiltonian as a generator
of $Sp(6,\mathcal{R})$. This allows the solution to the
operator Heisenberg equations of motion  to be expressed via
a matrix in $Sp(6,\mathcal{R})$, as also the time dependent
expectation values of the basic canonical variables.
Section~\ref{symp_trans} extends this approach to express
the relation between the variance matrices at two different
times as a symplectic congruence transformation. It is shown
that all the observable consequences of  the model  are
contained in such matrix relations. The consequences of the
$Sp(6,\mathcal{R})$ covariant statement of the Uncertainty
Principle for general states of the combined system, as well
as for special states of product form, are analyzed. The
relevance of the Williamson normal forms of variance
matrices is brought out. Section~\ref{concluding} contains
some concluding remarks.  
\section{The basic features of the $A$-$K$ model}
\label{ak_model}
The kinematic structure of the $A$-$K$ model is given by three
hermitian Cartesian position--momentum operator pairs: $\hat{q}$ and
$\hat{p}$ for system $\mathcal{S}$; and $\hat{Q}_1, \hat{P}_1,
\hat{Q}_2, \hat{P}_2$ for apparatus $\mathcal{A}$. We denote these
operators collectively by $\hat{\xi}_a, a=1, 2, \ldots, 6$:
\begin{equation}
\hat{\xi}_1=\hat{q},\, \hat{\xi}_2=\hat{p},
\,\hat{\xi}_3=\hat{Q}_1,\,
\hat{\xi}_4=\hat{P}_1,\, \hat{\xi}_5=\hat{Q}_2,\, \hat{\xi}_6=\hat{P}_2.
\label{operators}
\end{equation}
The canonical commutation relations (CCR) are:
\begin{equation}
[\hat{\xi}_a, \hat{\xi}_b]=i\hbar \beta_{ab}, \quad \quad
\beta=\left(\begin{array}{ccc}i\sigma_2 & 0 &0\\ 0& i\sigma_2& 0\\ 0
& 0 & i\sigma_2\end{array}\right).
\label{CCRs}
\end{equation} 
The natural covariance group of these CCR's is the
noncompact real symplectic group $Sp(6,\mathcal{R})$. This
will be defined  and exploited in succeeding Sections.
More  details about this group and
$Sp(2n,\mathcal{R})$ can be found
in~\cite{arvind_pramana}.

The measurement is described using the $A$-$K$ coupling of
$\mathcal{S}$ and $\mathcal{A}$ with the interaction Hamiltonian
\begin{equation}
\hat{H}=K_1\hat{q}\hat{P}_1 +K_2\hat{p}\hat{P}_2.
\label{interaction_hamiltonian}
\end{equation}
Which can be considered as the total Hamiltonian if we
neglect the `free' Hamiltonians for $\mathcal{S}$
and for $\mathcal{A}$ separately. Here we allow independent
choices of the real coupling constants $K_1$ and $K_2$ in
general which are of course of different physical
dimensions. This structure for $\hat{H}$ corresponds to
$\hat{Q}_1$ being the pointer position for measuring
$\hat{q}$, and $\hat{Q}_2$ for measuring $\hat{p}$.  Quantum
mechanics permits simultaneous measurements of $\hat{Q}_1$
and $\hat{Q}_2$, but not of $\hat{q}$ and $\hat{p}$. The aim
is to learn as much as quantum mechanics allows about the
latter from the former, based on the measurement
interaction~(\ref{interaction_hamiltonian}).

A general pure state Schr\"odinger wave function for $\mathcal{S}\
\oplus\ \mathcal{A}$ is written as $\Psi(q; Q_1, Q_2)$ with squared
norm
\begin{equation}
\|\Psi\|^2=\int_{-\infty}^\infty dq
\int_{-\infty}^\infty dQ_1\int_{-\infty}^\infty dQ_2
|\Psi(q; Q_1, Q_2)|^2.
\label{sq_norm}
\end{equation} 
The solution of the time dependent Schr\"odinger equation
\begin{equation}
i\hbar \frac{\partial}{\partial t}\Psi(q;
Q_1, Q_2; t)= \hat{H}\Psi(q; Q_1, Q_2; t)
\label{tdse}
\end{equation}
is easily obtained, using for instance the fact that $\hat{P}_1$ and
$\hat{P}_2$ are both constants of motion. The result is
\begin{eqnarray}
&&\Psi(q; Q_1, Q_2; t)=
\frac{1}{2\pi \hbar} \int_{-\infty}^\infty
\int_{-\infty}^\infty dP_2 dQ'_2\ {\rm
e}^{\frac{i}{\hbar}(Q_2-Q'_2)P_2}\nonumber\\ 
&&\quad\quad\times
\Psi(q-K_2tP_2; Q_1-K_1t(q-K_2\frac{t}{2}P_2), Q'_2;
0).
\label{tdse_solution}
\end{eqnarray} 
At any time $t>0$, a joint measurement of $\hat{Q}_1$ and
$\hat{Q}_2$ (permitted by quantum mechanics) yields results
${Q}_1, {Q}_2$ with the joint probability distribution
$P({Q}_1, {Q}_2; t)$ determined by the Born Rule:
\begin{equation} P({Q}_1, {Q}_2; t) =\int_{-\infty}^\infty
dq |\Psi(q; Q_1, Q_2; t)|^2.\end{equation} For a
general initial $\Psi(q; Q_1, Q_2; 0)$ this expression
cannot be simplified in any significant manner. In case
however the initial wave function for $\mathcal{S}\ \oplus\
\mathcal{A}$ is the product of a general pure state wave
function $\psi$ for $\mathcal{S}$ and two independent real
centered Gaussians for $\mathcal{A}$:
\begin{eqnarray}
&&\Psi(q; Q_1, Q_2;
0)=\sqrt{\frac{2}{\pi}}\frac{1}{(b_1b_2)^{1/4}}
\psi(q)\ e^{-Q_1^2/b_1-Q_2^2/b_2},
\nonumber\\
&&\quad\quad b_1,
b_2>0, \quad\quad
\int_{-\infty}^\infty dq\ |\psi(q)|^2=1,
\label{product_form}
\end{eqnarray}
Eq.~(\ref{tdse_solution}) simplifies to
\begin{eqnarray}
&&
\Psi(q; Q_1, Q_2;
t)=\frac{1}{\sqrt{2}} \frac{1}{\pi\hbar}
\left(\frac{b_2}{b_1}\right)^{1/4}\nonumber \\
&&\quad\quad\times \int_{-\infty}^\infty dP_2\ e^{iQ_2P_2/ \hbar}
\psi(q-K_2tP_2)\nonumber \\
&&\quad\quad\times
e^{-(Q_1-K_1tq+K_1K_2\frac{t^2}{2}P_2)^2/b_1-b_2P_2^2/4\hbar^2}.
\label{tdse_solution_simplified}
\end{eqnarray}
The probability distribution $P(Q_1, Q_2; t)$ also simplifies to
some extent and, after some algebra, takes the form:
\begin{eqnarray}
&&P(Q_1, Q_2; t)=\frac{1}{2}\frac{1}{(\pi\hbar K_2t)^2}
\sqrt{\frac{b_2}{b_1}}
\int_{-\infty}^\infty dq\ e^{-(Q_1-K_1tq)^2/b_1}\times\nonumber\\
&&\left|\int_{-\infty}^\infty dq' e^{-iQ_2q'/\hbar
K_2t}\psi(q')
e^{-X(t)(q-q')^2-(Q_1-K_1tq')^2/2b_1}\right|^2,\nonumber\\
&&\quad\quad{\rm with}\quad X(t)=\frac{b_2}{(2\hbar
K_2t)^2}-\frac{K_1^2t^2}{4b_1}.
\label{simplified_pd}
\end{eqnarray}

In the original $A$-$K$ model, and in some later work as
well, some more simplifying assumptions were made. In
suitable units, including $\hbar=1$, (i) $K_1=K_2=K$; (ii)
$b_1=b_2^{-1}=b$; (iii) $Kt=1$. In this situation,
Eq.~(\ref{simplified_pd}) simplifies one more step, since
$X(t)=0$ and the $q$ integration can be carried out:
\begin{equation}
P(Q_1, Q_2; t)=\frac{1}{\sqrt{2 \pi^2 b}}
 \left|\int_{-\infty}^\infty dq' \ \psi(q')\ 
e^{-iq'Q_2- (q'-Q_1)^2/2b}\right|^2.
\label{simple_simple_pd}
\end{equation} 
We will however work with the more general
expression~(\ref{simplified_pd}), especially since the
simplifying assumptions mentioned conflict with $K_1$ and
$K_2$ being of different physical dimensions, while $b_1$
and $b_2$ are both squared lengths.
\section{The group $Sp(6,\mathcal{R})$ and its uses}
\label{sp6r}
The CCR's~(\ref{CCRs}) are preserved by real linear
canonical transformations acting on $\hat{\xi}_a$. These
form the 21-parameter noncompact real symplectic group in 6
real dimensions defined by
\begin{equation}
\!\!\!Sp(6,\mathcal{R})\!=\!\{S\!=\!(S_{ab})\!=\!{\rm real}\,
6\!\times\! 6\, {\rm
matrix}\ |S^T\beta S=\beta\},\!
\label{sp6r_group}
\end{equation}
where $\beta$ is the symplectic metric matrix given in
Eq.~(\ref{CCRs}).
Thus we have:
\begin{eqnarray}
&& S\in Sp(6,\mathcal{R}),\,\,
\hat{\xi}'_a=S_{ba}\hat{\xi}_b\Rightarrow [\hat{\xi}'_a,
\hat{\xi}'_b]=i\hbar \beta_{ab},\nonumber\\
&&\hat{\xi}'_a=\mathcal{U}(S)\hat{\xi}_a \mathcal{U}(S)^{-1},
\end{eqnarray}
where the unitary operators $\mathcal{U}(S)$ are determined up to signs
by $S$ and obey
\begin{equation}
\mathcal{U}(S') 
\mathcal{U}(S)=\pm 
\mathcal{U}(S'S),
\end{equation}
thus constituting the two-valued metaplectic unitary
representation of
$Sp(6,\mathcal{R})$~\cite{arvind_pramana,arvind_pra94}.
Since $\hat{\xi}_{1, 3, 5}$ are lengths and $\hat{\xi}_{2,
4, 6}$ are momenta, each element $S_{ba}$ of $S$ carries a
suitable physical dimension. The $\mathcal{U}(S)$ are
(products of) exponentials of antihermitian quadratic
expressions in the $\hat{\xi}_a$.

The relevance of $Sp(6,\mathcal{R})$ arises from the fact
that the Hamiltonian~(\ref{interaction_hamiltonian}) is quadratic in the
$\hat{\xi}$'s. We have
\begin{eqnarray}
\hat{H}=\frac{1}{2}h_{ab}\hat{\xi}_a\hat{\xi}_b, \quad
h_{ab}=h_{ba},\nonumber\\
h=\left(\begin{array}{cccccc}
0 & 0 & 0 & K_1 & 0 & 0\\
0 & 0 & 0 & 0 & 0 & K_2\\
0 & 0 & 0 & 0 & 0 & 0\\
K_1 & 0 & 0 & 0 & 0 & 0\\
0 & 0 & 0 & 0 & 0 & 0\\
0 & K_2 & 0 & 0 & 0 & 0\\
\end{array}\right),
\label{hamiltonian_symplectic}
\end{eqnarray}
the only nonzero elements of $h$ being $h_{14}=h_{41}=K_1$
and $h_{26}=h_{62}=K_2$.
Therefore we get
\begin{eqnarray}
&&[\hat{H}, \hat{\xi}_a]=i\hbar
J_{ba}\hat{\xi}_b,\quad  [\hat{H},  [\hat{H},
\hat{\xi}_a]]=-\hbar^2 (J^2)_{ba}\hat{\xi}_b,
\quad\ldots,\nonumber\\[6pt]
&& J=h\beta=\left(\begin{array}{cccccc}
0 & 0 & -K_1 & 0 & 0 & 0\\
0 & 0 & 0 & 0 & -K_2  & 0\\
0 & 0 & 0 & 0 & 0 & 0\\
0 & K_1 & 0 & 0 & 0 & 0\\
0 & 0 & 0 & 0 & 0 & 0\\
-K_2 & 0 & 0 & 0 & 0 & 0\\
\end{array}\right),\nonumber\\
&&J^2=K_1K_2\left(\begin{array}{cccccc}
0 & 0 & 0 & 0 & 0 & 0\\
0 & 0 & 0 & 0 & 0  & 0\\
0 & 0 & 0 & 0 & 0 & 0\\
0 & 0 & 0 & 0 & -1 & 0\\
0 & 0 & 0 & 0 & 0 & 0\\
0 & 0 & 1 & 0 & 0 & 0\\
\end{array}\right),\quad J^3=0.
\end{eqnarray}

This real matrix $J$ obeys
\begin{equation}
J^T\beta+\beta J=0,\quad {\it i.e.}\quad (\beta
J)^J=\beta J.\end{equation} As a result, the Heisenberg
equations of motion for $\hat{\xi}_a(t)$ are explicitly
solvable, containing only three terms:
\begin{eqnarray}
\hat{\xi}_a(t)&=&e^{i\hat{H}t/\hbar}\hat{\xi}_a
e^{-i\hat{H}t/\hbar}\nonumber\\
&=&\hat{\xi}_a-t
J_{ba}\hat{\xi}_b+\frac{t^2}{2}(J^2)_{ba}\hat{\xi}_b\nonumber\\
&=&(S(t)^T\hat{\xi})_a,\nonumber\\
S(t)&=&e^{-tJ} = \left(\begin{array}{cccccc}
1 & 0 & tK_1 & 0 & 0 & 0\\
0 & 1 & 0 & 0 & tK_2  & 0\\
0 & 0 & 1 & 0 & 0 & 0\\
0 & -tK_1 & 0 & 1 & -\frac{t^2}{2}K_1K_2 & 0\\
0 & 0 & 0 & 0 & 1 & 0\\
tK_2 & 0 & \frac{t^2}{2}K_1K_2 & 0 & 0 & 1
\end{array}\right)\nonumber  \\[12pt]
&&
\quad \quad \quad\quad 
\quad \quad \quad\quad 
\in Sp(6,\mathcal{R})
\label{symp_solution}
\end{eqnarray}
For any solution $\vert\Psi(t)\rangle$ of the time dependent Schr\"odinger
equation~(\ref{tdse}), the symplectic matrix $S(t)$ determines the
evolution of the expectation values of the operators $\hat{\xi}_a$:
\begin{eqnarray}
{\xi}_a(t)&=&\langle \Psi(t)| \hat{\xi}_a|\Psi(t)\rangle\nonumber\\
 &=&\langle \Psi(0)| \hat{\xi}_a(t)|\Psi(0)\rangle\nonumber\\
 &=&S(t)_{ba}\xi_b(0).
\end{eqnarray} 
For $a=3$ and $5$ we get:
\begin{eqnarray}
Q_1(t)=S(t)_{b3}\xi_b(0)= Q_1(0)+tK_1q(0)+\frac{t^2}{2}K_1K_2P_2(0),\nonumber\\
Q_2(t)=S(t)_{b5}\xi_b(0)=
Q_2(0)+tK_2p(0)-\frac{t^2}{2}K_1K_2P_1(0). \nonumber \\
\label{Q1Q2_equations}
\end{eqnarray} 
Here, $q(0)$ and $p(0)$ are the quantum mechanical
expectation values of $\hat{q}$ and $\hat{p}$ of
$\mathcal{S}$ in the state $\vert \Psi(0)\rangle$, obtained from
independent measurements of these operators one at a time.
The relations between the expectation values $Q_1(t),
Q_2(t)$ of $\hat{Q}_1, \hat{Q}_2$ in $\vert\Psi(t)\rangle$ and
$q(0), p(0)$ are encoded in the matrix $S(t)$.  The `pointer
readings' $Q_1(t), Q_2(t)$ reveal the properties $q(0),
p(0)$ of $\mathcal{S}$ consistent with $[\hat{Q}_1,
\hat{Q}_2]=0, [\hat{q}, \hat{p}]\ne 0$.

In case $\vert\Psi(0)\rangle$ is of the product
form~(\ref{product_form}), Eqs.~(\ref{Q1Q2_equations}) simplify to 
\begin{equation}Q_1(t)=tK_1q(0),\quad
Q_2(t)=tK_2p(0).
\label{Q1Q2_solution_simple}
\end{equation}
\section{Symplectic transformation law for variance matrices}
\label{symp_trans}
We first recall the $Sp(6,\mathcal{R})$ covariant form of the Uncertainty
Principle for general states of the composite system $\mathcal{S}\
\oplus\ \mathcal{A}$. The state is in general a mixed one described
by a density matrix $\hat{\rho}$; in the pure case
$\hat{\rho}=|\Psi\rangle\langle \Psi|$. The 6 dimensional real
symmetric positive definite variance matrix $V=(V_{ab})$ is defined
by
\begin{eqnarray}
V_{ab}&=&{\rm Tr}[\hat{\rho}\frac{1}{2}\{\hat{\xi}_a,
\hat{\xi}_b\}]-\langle \hat{\xi}_a\rangle \langle
\hat{\xi}_b\rangle,\nonumber\\ 
\langle\hat{\xi}_a\rangle
&=&Tr(\hat{\rho}\hat{\xi}_a).
\label{variance_matrix}
\end{eqnarray}
Under $Sp(6,\mathcal{R})$ action on
$\hat{\rho}$, the effect on $V$ is a matrix congruence
transformation:
\begin{equation}
S\in Sp(6,\mathcal{R}): \,
\hat{\rho}'=\mathcal{U}(S)^{-1}\hat{\rho}\mathcal{U}(S)\Rightarrow
V'=S^TVS.
\label{variance_transform}
\end{equation}
The statement of the Uncertainty Principle in $Sp(6,\mathcal{R})$ covariant
form is
\begin{equation}
V+\frac{i}{2}\hbar \beta\ge 0.
\end{equation}
Every matrix $V$ obeying this condition is realizable (in general in
infinitely many ways) as the variance matrix of some physical state.
In particular, if $V$ is physically realizable, then so is $S^TVS$
for any $S\in Sp(6,\mathcal{R})$~\cite{simon_nm_sud}.

Ignoring for the moment the physical dimensions of each $V_{ab}$ and
each $S_{ab}$, a given numerical matrix $V$ which is real symmetric
positive definite and obeys $V+\frac{i}{2}\beta\ge 0$ can be written
in the form
\begin{eqnarray}
&& V=S^T_0\ V_d\ S_0, \quad S_0\in Sp(6,\mathcal{R}),\nonumber\\
&& V_d={\rm diag}(\kappa_1, \kappa_1, \kappa_2, \kappa_2,
\kappa_3, \kappa_3),\nonumber\\
&& \kappa_j\ge \frac{1}{2},\quad j=1, 2, 3.
\label{williamson}
\end{eqnarray}
This is the Williamson normal form of $V$ and here $S_0$ too is
purely numerical. For our problem with each $V_{ab}, S_{ab}$
carrying definite dimensions, we need a modified normal form which
is:
\begin{eqnarray}
V&=&S_0^{'T}V'_dS'_0, \quad S'_0\in
Sp(6,\mathcal{R}),\nonumber\\
 V'_d&=&{\rm diag}(\kappa_1, \kappa'_1, \kappa_2, \kappa'_2,
\kappa_3, \kappa'_3), \quad \kappa_j,
\kappa'_j>0,\nonumber\\
&& \kappa_j\kappa'_j\ge \frac{\hbar^2}{4},\quad j=1, 2, 3.
\label{williamson_dim}
\end{eqnarray}
Now each $\kappa_j$ is a squared length and each $\kappa'_j$ a
squared momentum. The passage from\~(\ref{williamson})
to~(\ref{williamson_dim})  involves
reciprocal scale transformations within each canonical pair, and
these are elements of $Sp(6,\mathcal{R})$.

Now we apply these results to the $A$-$K$ model. For simplicity we
limit ourselves to pure states $\vert\Psi(0)\rangle $ at $t=0$, though
this is not essential. Then the variance matrix $V(t)$ evolves via
$S(t)\in Sp(6,\mathcal{R})$ in Eq.~(\ref{symp_solution}):
\begin{eqnarray}
V_{ab}(t)&=&\langle \Psi(t)\vert\frac{1}{2}\{\hat{\xi}_a,
\hat{\xi}_b\}\vert\Psi(t)\rangle 
\nonumber \\ &&
-\langle \Psi(t)|
\hat{\xi}_a|\Psi(t)\rangle \langle\Psi(t)|\hat{\xi}_b
|\Psi(t)\rangle.\nonumber\\
V(t)&=&S(t)^TV(0)S(t),
\end{eqnarray}
Let us focus on the uncertainties or spreads in the pointer
positions, $(\Delta Q_1)^2=V_{33}$ and $(\Delta
Q_2)^2=V_{55}$. In the diagonal form of $V'_d$ in
Eq.~(\ref{williamson_dim}), we have $\Delta
Q_1=\kappa_2^{1/2}, \Delta Q_2=\kappa_3^{1/2}$ and there is
no lower bound on $\kappa_2\kappa_3$. Based on the general
$Sp(6,\mathcal{R})$ transformation
rule~(\ref{variance_transform}) for $V$, it is easy to see
that for any chosen $t>0$, we can always choose $V(0)$ so
that $(\Delta Q_1)(t)$ and $(\Delta Q_2)(t)$ are each as
small and positive as we wish.  Thus as expected, quantum
mechanics does not imply any universal lower bound for the
uncertainty product $\Delta Q_1\Delta Q_2$.

To study the time dependences of $(\Delta Q_1)(t)$ and
$(\Delta Q_2)(t)$  further, we need to evaluate $V_{33}(t),
V_{55}(t)$ respectively. In each case, since many elements
of $S(t)$ vanish, only six terms remain, three of direct
type and three cross terms:
\begin{eqnarray}
&&(\Delta Q_1)^2(t)= V_{33}(t)= S_{a3}(t) S_{b3}(t) V_{ab}(0)\nonumber\\
&&= V_{33}(0)+(tK_1)^2 V_{11}(0)+\frac{1}{4}(t^2K_1K_2)^2 V_{66}(0) \nonumber \\
&& +2tK_1 V_{13}(0)+t^2K_1K_2 V_{36}(0)+t^3K_1^2K_2 V_{16}(0) \nonumber \\
&& =(\Delta Q_1)^2(0)+(tK_1)^2 (\Delta
q)^2(0)+\frac{1}{4}(t^2K_1K_2)^2 (\Delta P_2)^2(0) \nonumber
\\
&& +3\ {\rm  cross\ terms};  
\label{varq1}
\\[12pt]
&& (\Delta Q_2)^2(t)= V_{55}(t)= S_{a5}(t) S_{b5}(t) V_{ab}(0) \nonumber \\
&& =(\Delta Q_2)^2(0)+(tK_2)^2 (\Delta
p)^2(0)+\frac{1}{4}(t^2K_1K_2)^2 (\Delta P_1)^2(0) \nonumber
\\
&& +3\ {\rm  cross\ terms}.
\label{varq2}
\end{eqnarray}
In comparison with Eq.~(\ref{Q1Q2_solution_simple}), we see
that just as the mean values $Q_1(t), Q_2(t)$ lead to the
mean values $q(0), p(0)$ of the operators $\hat{q}, \hat{p}$
of $\mathcal{S}$, now $(\Delta Q_1)(t), (\Delta Q_2)(t)$
yield the spreads $(\Delta q)(0), (\Delta p)(0)$ in
$\hat{q}, \hat{p}$ at $t=0$.

Let us finally consider $\vert\Psi(0)\rangle$ to be of the
product form Eq.(\ref{product_form}). The initial variance
matrix $V(0)$ is now block diagonal:
\begin{eqnarray}
&&V(0)=\left(\begin{array}{cccc}
\begin{array}{cc}V_{11}(0) & V_{12}(0)\\ V_{21}(0) &
V_{22}(0) \end{array} & 0 & 0\\0 &
\begin{array}{cc} \frac{1}{4}b_1 & 0\\ 0 &
\frac{\hbar^2}{b_1} \end{array} & 0\\
0 & 0 &
\begin{array}{cc} \frac{1}{4}b_2 & 0\\ 0 &
\frac{\hbar^2}{b_2} \end{array}
\end{array}\right),\nonumber\\[12pt]
&&\quad\quad\quad V_{11}(0)=(\Delta q)^2(0),\nonumber \\
&& \quad\quad\quad V_{22}(0)=(\Delta p)^2(0),\nonumber \\
&&\quad\quad\quad V_{12}(0)=(\Delta q \Delta p)(0).
\end{eqnarray}
All cross terms in Eqns.~(\ref{varq1})~\&~(\ref{varq2})  vanish, and we find:
\begin{eqnarray}
(\Delta Q_1)^2(t)= (tK_1)^2 (\Delta
q)^2(0)+\frac{1}{4b_2}(b_1b_2+(t^2\hbar K_1K_2)^2),\nonumber
\\
(\Delta Q_2)^2(t)= (tK_2)^2 (\Delta
p)^2(0)+\frac{1}{4b_1}(b_1b_2+(t^2\hbar K_1K_2)^2).
\nonumber \\ \end{eqnarray}
Using $\Delta q(0)\Delta p(0)\ge \hbar/2$, we find a lower bound for
$\Delta Q_1(t)\Delta Q_2(t)$:
\begin{eqnarray}
&&(\Delta Q_1(t) \Delta Q_2(t))^2= (t^2K_1K_2)^2 (\Delta
q(0)\Delta
p(0))^2
\nonumber \\
&&+\frac{1}{4}(b_1b_2+(t^2\hbar K_1K_2)^2)
\nonumber \\ &&\quad \times 
\left(\frac{(tK_1)^2}{b_1} (\Delta
q)^2(0)+\frac{(tK_2)^2}{b_2}(\Delta
p)^2(0)\right)
\nonumber \\
&&
+\frac{1}{16 b_1b_2}(b_1b_2+(t^2\hbar
K_1K_2)^2)^2\nonumber\\
&& \ge (b_1b_2+2\sqrt{b_1b_2}\hbar t^2K_1K_2+(t^2\hbar
K_1K_2)^2)^2/16b_1b_2,\nonumber \\
\label{q1q2variances}
\end{eqnarray} \ie ,
\begin{equation}
\Delta Q_1(t)\Delta Q_2(t)\ge
\frac{1}{4\sqrt{b_1b_2}}(\sqrt{b_1b_2}+ t^2\hbar
K_1K_2)^2.\end{equation}
However, as is to be expected, this `Uncertainty Principle'
for $\Delta Q_1(t) \Delta Q_2(t)$ depends both on the
parameters $K_1, K_2$ in the
Hamiltonian~(\ref{interaction_hamiltonian}) and on the
parameters $b_1, b_2$ in the initial product wave
function~(\ref{product_form}).

\section{Concluding Remarks}
\label{concluding}
In this paper we have shown the usefulness of methods based
on the symplectic group $Sp(6,\mathcal{R})$ for analyzing the
consequences of the Arthurs-Kelly model of measurement in
quantum mechanics. All physically meaningful results are
expressed in the connections  between time dependent
variance matrices at different times. These connections are
stated in terms of specific real six dimensional  symplectic
matrices.  The variance matrices are made up of averages and
spreads of one observable at a time. So even though the
model is an attempt to give meaning to simultaneous
measurements of position and momentum in quantum mechanics,
only the variance matrices are involved in the  statement of
observable consequences  of the model. The presence of two
independent coupling constants in the interaction
Hamiltonian allows us to study various options of
measurement sequences with some flexibility.

From the
Eqns.~(\ref{Q1Q2_solution_simple})~\&~(\ref{q1q2variances})
it is clear that, a measurement of commuting variables
$\hat{Q}_1$ and $\hat{Q}_2$ at time $t$, reveals information
about non-commuting variable $\hat{q}$ and $\hat{p}$ at
$t=0$.  If these measurements are repeated several times,
the mean values and the variances at time $t=0$ of $\hat{q}$
and $\hat{p}$ can be estimated. These estimates will contain
additional noise because we are estimating non-commuting
observables. These additional noise terms are the last terms
in Eqns.~(\ref{q1q2variances}). The strength of the
$\hat{q}$ measurement is determined by $t K_1$ and $b_1$.
The larger the values of $t K_1$ the stronger is the
measurement and the smaller the value of $b_1$,
the stronger is the measurement. In a similar manner, the
strength of the $\hat{p}$ measurement is dictated by the
values of the parameters $t K_2$ and $b_2$.

The formalism presented here allows us to explore different
regimes, for instance, we can easily switch off one
measurement by setting $K_1$ or $K_2$ to zero. We can
perform sequential measurements  by first having a non-zero
$K_1$ with $K_2$ set to zero and vice versa. In such a
sequential measurement,  the first measurement is required
to be weak, in the sense that $t K_1$ is small or $b_1$ is
large.  The second measurement could  be a strong one and
close to a projective measurement.

We can try to estimate the state of the system from the
measurements of $\hat{Q}_1$ and $\hat{Q}_2$. Such estimates
can be done in many ways: independent measurements,
sequential measurements, and joint measurements. What we would
like to emphasize is that symplectic techniques are useful
in all these scenarios. Some of these aspects will be taken
up elsewhere. 
\begin{acknowledgments}
One of the authors (NM) thanks the Indian National
Science Academy for award of INSA Distinguished
Professorship during whose tenure this work was initiated.
\end{acknowledgments}
%
\end{document}